\documentclass[aps,pra,reprint,superscriptaddress,longbibliography]{revtex4-2}
\usepackage[utf8]{inputenc}
\usepackage[english]{babel}
\usepackage[T1]{fontenc}
\usepackage{amssymb}
\usepackage{amsmath}
\usepackage{amsfonts} 
\usepackage{amsthm}
\usepackage{mathtools}
\usepackage{hyperref}
\usepackage{float}
\usepackage{enumitem}
\usepackage{bbold}
\usepackage{natbib}
\usepackage{orcidlink}
\usepackage{algorithm}
\usepackage{algpseudocode}
\usepackage{appendix}
\usepackage[normalem]{ulem}

\newtheorem{result}{Result}

\begin{document}

\title{Hamiltonian simulation with explicit formulas for Digital-Analog Quantum Computing}

\author{Mikel \surname{Garcia de Andoin}
\orcidlink{0000-0002-5009-7109}}\email{mikel.garciadeandoin@ehu.eus}\affiliation{Department of Physical Chemistry, University of the Basque Country UPV/EHU, Apartado 644, 48940 Leioa, Spain}
\affiliation{EHU Quantum Center, University of the Basque Country UPV/EHU, Barrio Sarriena s/n, 48940 Leioa, Spain}

\author{Thorge Müller\orcidlink{0009-0000-9561-5945}}
\email{thorge.mueller@dlr.de}
\affiliation{%
 Department High-Performance Computing, Institute of Software Technology, German Aerospace Center (DLR), 51147 Cologne, Germany
}%

\author{Gonzalo Camacho\orcidlink{0000-0001-6900-8850}}\email{gonzalo.camacho@dlr.de}
\affiliation{%
 Department High-Performance Computing, Institute of Software Technology, German Aerospace Center (DLR), 51147 Cologne, Germany
}%

\begin{abstract}
    Digital-analog is a quantum computational paradigm that employs the natural interaction Hamiltonian of a system as the entangling resource, combined with single qubit gates, to implement universal quantum operations.
    As in the case of its digital gate-based counterpart, designing digital-analog circuits that employ optimal quantum resources often requires an exceedingly large classical computational time. In this work we find a suboptimal solution to this exponentially large problem, showing that it can be solved within polynomial computational time. In particular, we provide an exact solution for the problem of expressing arbitrary two-body Hamiltonians as the sum of local unitary transformations of an arbitrary Ising Hamiltonian, with the total number of required terms being at most quadratic in system size. 
    This allows us to design a digital-analog simulation protocol that avoids employing numerical optimization over a large parameter space at the preprocessing stage, minimizing computational resources and allowing for further scaling.
\end{abstract}

\maketitle

\section{Introduction}\label{sec:intro}

Universal quantum computation can be achieved through the application of different operations. In digital quantum computing one usually employs a universal set of discrete gates composed of arbitrary single qubit gates (SQG) and two-qubit gates~\cite{Lloyd1995Univesal, Barenco1995Elementary}. On the contrary, digital-analog quantum computing (DAQC) employs the natural interaction Hamiltonian of a given system as the entangling resource~\cite{Adrian2020DAQC}. By additionally implementing arbitrary SQGs on top of this evolution, we can implement any unitary operation~\cite{Dodd2002UnivQC}. This paradigm allows us to access the universality of the digital paradigm while benefiting from the noise resilience of analog quantum computations~\cite{Paula2024}. 

Although these set of operations are universal, composing the circuit that implements a particular operation employing the minimum resources is a highly non-trivial question~\cite{Moro2021,Haferkamp2022,van2025optimal}. When compiling a quantum circuit, minimizing the total number of gates employed and the total circuit time are essential computational efficiency factors~\cite{YanJunchi2024review}. In digital quantum computing, the total circuit time is dominated by the circuit depth and the total number of two-qubit gates, given that the application times of two-qubit gates are often slower than those of SQGs~\cite{Nori2012,Joel2023,Shapira2023,chen2025efficient}. In DAQC, the total time is also dominated by the time where the system has been let to evolve freely under the natural Hamiltonian of the system, where all entangling operations are encoded~\cite{Galicia2020,Bassler2024timeoptimalmulti}.

In both paradigms, finding the optimal circuit is considered to be an NP-Hard problem~\cite{Bassler2023Synthesis, MikelAlvaro2024, kjelstrom2025exactquantumcircuitoptimization}, and thus, it is expected to consume exponentially many computing resources. As we do not expect this task to be efficiently solvable, one usually resorts to algorithms that take reduced time to find a suboptimal solution. While these protocols are usually available for the digital paradigm \cite{Amy2014,Ruiz2025,yan2025}, they remain outstanding for digital-analog. The available methods to date require exponential resources to synthesize an optimal DAQC circuit for solving problems involving arbitrary Hamiltonians~\cite{MikelAlvaro2024,bassler2025general}. Solving these problems is fundamental not only for optimal generation of circuits in the DAQC paradigm, but also for the field of quantum simulation~\cite{Lloyd1996,Buluta2009,Cirac2012}, where impressive experimental advances have been achieved in the last years~\cite{Barreiro2011,Blatt2012,Bloch2012,Gross2017,Bernien2017,Choi2020,Ebadi2021}. Indeed, it is expected that DAQC will be able reproduce these milestones in quantum computing platforms such as superconducting circuits~\cite{Lamata2018,abughanem2025superconducting}, trapped ions~\cite{arrazola2016digital,monroe2021}, or neutral atoms~\cite{Lu_2024,bluvstein2024logical}, as these setups already provide all the capabilities required for the practical implementation of digital-analog circuits. Successfully simulating quantum systems might have direct applications in the fields of chemistry or condensed matter simulations among others \cite{Georgescu2014,McArdle2020,clinton2024towards}. 

\begin{figure}[t]
    \centering
    \includegraphics[width=\linewidth]{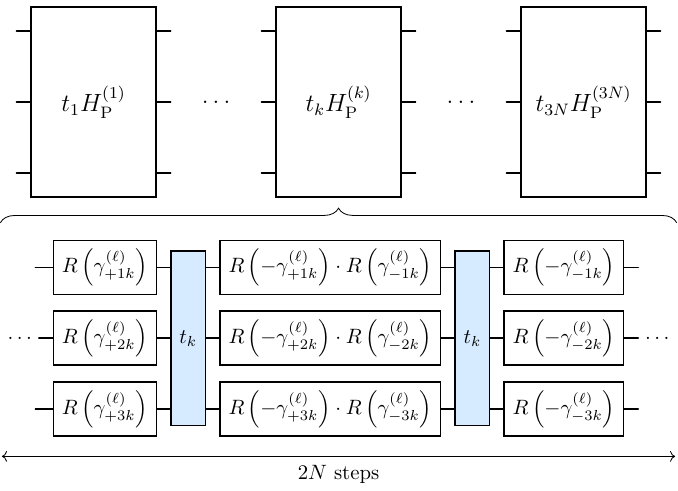}
    \caption{Problem Hamiltonian decomposition and DAQC circuit sketch. The colored blocks corresponds to analog blocks with evolution time $t_k$. The full circuit that simulates the evolution under the problem Hamiltonian for a time $T$, $TH_\text{P}=\sum_k t_kH_\text{P}^{(k)}$, is composed of up to $12N^2$ digital-analog blocks. The particular values of the rotation angles of the SQGs and $t_k$ are given in Eq.~\eqref{eq:fulldecomposition}.}
    \label{fig:protocolsketch}
\end{figure}

In this work, we introduce a protocol for obtaining a suboptimal compilation in the simulation of arbitrary two-body Hamiltonians employing a DAQC circuit (See Fig.\ref{fig:protocolsketch}). Instead of restricting the protocol to single qubit gates from the Clifford group~\cite{Bassler2023Synthesis, MikelAlvaro2024, kjelstrom2025exactquantumcircuitoptimization}, here we employ arbitrary single qubit gates to design the circuit. The proposed protocol relies on the eigendecomposition a couplings matrix having size $3N\times3N$, with $N$ being the number of qubits in the system. This allows us to produce a valid DAQC circuit with minimal computational resources, in particular in time $\mathcal{O}(N^3)$. The protocol relies on closed-form formulas, thus removing the need to employ classical optimization algorithms to design the circuit. The resulting circuit is composed of at most $12N^2$ digital-analog blocks, which has a constant factor overhead over previous protocols, which required up to $9N(N-1)/2$ blocks. This way, we maintain the same asymptotic scaling as the state-of-the-art optimal protocols.

The rest of the paper is structured as follows. In Sec.~\ref{sec:daqc}, we introduce the DAQC paradigm in detail and point out the main problem we are aiming to solve. Sec.~\ref{sec:result} introduces the main result of the paper providing a step-by-step description. In Sec.~\ref{sec:numerics} we report numerical calculations comparing the output circuit times with previously known results. Finally, in Sec.~\ref{sec:conclusion} we discuss the results and provide some perspectives on the potential future applications of the protocol. 

\section{Preliminaries on Digital Analog Quantum Computing}\label{sec:daqc}

We can always describe a unitary operation in terms of the consecutive application of single qubit gates (SQGs) and evolution under entangling two-body Hamiltonians~\cite{Dodd2002UnivQC}. In DAQC these operations are implemented through the application of digital blocks, composed of SQGs, and analog blocks, in which the system freely evolves under the natural Hamiltonian of the system, the source Hamiltonian $H_{\text{S}}$~\cite{Adrian2020DAQC}. However, connecting the unitary evolution of a desired problem Hamiltonian $H_{\text{P}}$ and our resources is a nontrivial task in DAQC. 

As a simple example, let us show how to simulate the evolution under a time-independent two-body ZZ Hamiltonian
\begin{equation}\label{eq:problem_zz}
    H_\text{P}^{zz}=\sum_{i<j}{h_\text{P}}_{ij}^{zz}\sigma_i^z\sigma_j^z,
\end{equation}
during a time $T$ using as a resource another ZZ Hamiltonian 
\begin{equation}\label{eq:zz_hamiltonians}    
    H_\text{S}=\sum_{i<j}{h_\text{S}}_{ij}^{zz}\sigma_i^z\sigma_j^z.
\end{equation}
The $\sigma_i^{\alpha=x,y,z}$ represent the standard Pauli matrices acting on qubit $i$, whereas ${h_\text{P}}_{ij}^{zz},{h_\text{S}}_{ij}^{zz}$ represent the couplings in the problem and source Hamiltonians, respectively. This Hamiltonian is characteristic of trapped ions with Magnetic Gradient Induced Couplings (MAGIC)~\cite{monroe2021}, among other systems. We employ a DAQC protocol in which the SGQs are restricted to be $X\coloneqq \sigma^x$ gates~\cite{Adrian2020DAQC}. This choice allows us to flip the effective sign of a coupling by conjugating it with an $X$ gate in one of the two qubits, $\sigma_i^x e^{-iT\sigma_i^z\sigma_j^z}\sigma_i^x=e^{+iT\sigma_i^z\sigma_j^z}$. With this, the problem is reduced to solving
\begin{equation}\label{eq:DAQCcircuit}
    e^{-iTH_\text{P}^{zz}}=\prod_k U_k e^{-it_kH_\text{S}}U_k^\dagger=\prod_k e^{-it_kH_\text{S}^{(k)}},
\end{equation}
with $U_k$ the digital block at step $k$ containing the $X$ gates, and $H_\text{S}^{(k)}$ is the effective source Hamiltonian at step $k$. In this case, all effective Hamiltonians commute with each other, $[H_\text{S}^{(k)},H_\text{S}^{(k')}]=0\ \forall k,k'$, and thus the equation can be solved exactly equating the exponents,
\begin{equation}\label{eq:DAQCexponents}
\begin{split}
    T\sum_{i<j}{h_\text{P}}_{ij}^{zz}\sigma_i^z\sigma_j^z &= \sum_k t_k H_\text{S}^{(k)}\\
    &=\sum_k t_k\sum_{i<j}\gamma_{ik}\gamma_{jk}{h_\text{S}}_{ij}^{zz}\sigma_i^z\sigma_j^z,
\end{split}
\end{equation}
where $\gamma_{ik}=-1$ if an $X$ is applied to the $i$-th qubit at the $k$-th step and $\gamma_{ik}=1$ otherwise. Here, the unknown parameters are the analog block times $t_k$, as the effective Hamiltonians are fixed by the election of the digital blocks. 

We can further rearrange this equation equating both sides term by term. We therefore employ a notation in which we express an arbitrary two-body Hamiltonian, $H=\sum_{i< j}\sum_{\mu,\nu}h_{ij}^{\mu\nu}\sigma_i^\mu\sigma_j^\nu$, by arranging its coupling terms into a matrix $\mathbf{H}$, $\mathbf{H}_{3i+\mu,3j+\nu}=h_{ij}^{\mu\nu}$, with every qubit index $i$ starting at $i=0$. By using this notation, and by rearranging the Hamiltonian parameters to the right hand side of the equation, we end up with a simple linear system of equations of the form
\begin{equation}\label{eq:DAQCcompact}
    Mt=T \mathbf{H}_\text{P}\oslash\mathbf{H}_\text{S},
\end{equation}
where $M$ is a tensor representing the changes in the effective Hamiltonians at each step, $t$ is the vector containing the analog block times and $\oslash$ is the element-wise or Hadamard division. The main goal is to reduce as much as possible the total time to implement the simulation, $t_\text{A}\sim \lVert t\rVert_1$. As shown in Ref.~\cite{MikelAlvaro2024}, solving this problem exactly might be exponentially costly, although greedy algorithms tend to provide a solution in polynomial time. 

In general, we do not want to restrict ourselves to Hamiltonians of the form given by Eq.~\eqref{eq:zz_hamiltonians}. Instead, we want to be able to perform the simulation for an arbitrary problem using any given two-body Hamiltonian as a resource. 

As a starting point, let us assume a situation in which ${h_\text{S}}_{ij}^{\mu\nu}\neq0$ if ${h_\text{P}}_{ij}^{\mu\nu}\neq0$. In Ref.~\cite{MikelAlvaro2024}, the proposed DAQC schedules contained digital blocks with all possible combinations of the Pauli gates. This way, the system of equations can be written as
\begin{equation}\label{eq:arbitraryDAQCexponents}
    T\sum_{i<j}\sum_{\mu,\nu}{h_\text{P}}_{ij}^{\mu\nu} \sigma_i^\mu\sigma_j^\nu = \sum_k t_k\sum_{i<j}\sum_{\mu,\nu}\gamma_{ik}^{\mu}\gamma_{jk}^{\nu}{h_\text{S}}_{ij}^{\mu\nu}\sigma_i^\mu\sigma_j^\nu,
\end{equation}
where $\mu\nu$ takes the indices of the Pauli gates, $\{\mu,\nu\}\in\{x,y,z\}$. For this initial proposal, $\gamma_{ik}^{\mu}$ could only take values of $\pm1$. However, a later proposal designed DAQC circuit in which the digital blocks contained combinations of the SQGs in the Clifford group, allowing to lift the requirement for solving the equations when ${h_\text{S}}_{ij}^{\mu\nu}\neq0$ for at least one combination of $\{\mu,\nu\}$ if ${h_\text{P}}_{ij}^{\mu\nu}\neq0$ for any $\{\mu,\nu\}$~\cite{Bassler2023Synthesis}. In this case, the most general system of equations can be written as:
\begin{equation}\label{eq:daqc_sand2}
    T\sum_{i<j}\sum_{\mu,\nu}{h_\text{P}}_{ij}^{\mu\nu} \sigma_i^\mu\sigma_j^\nu\\
    =\sum_k t_k\sum_{i<j}\sum_{\mu,\nu,\xi,\eta}\gamma_{ik}^{\xi\mu}\gamma_{jk}^{\eta\nu}{h_\text{S}}_{ij}^{\xi\eta}\sigma_i^\mu\sigma_j^\nu,
\end{equation}
with $\{\xi,\eta\}\in\{x,y,z\}$. However, as the set of gates employed is discrete, we can write these equations in a similar fashion as in Eq.~\eqref{eq:arbitraryDAQCexponents} by including the information of the extra components $\{\xi,\eta\}$ in the definition of each block $k$. There is an additional proposal in Ref.~\cite{MikelAlvaro2024} in which arbitrary SQGs can be applied during the digital blocks. This also relaxes the requirements on $H_\text{S}$, but it has a drawback. The computational cost of obtaining the optimal SQGs for the circuit is itself a demanding task, as this proposal required constructing an matrix product state (MPS) proxy for evaluating the DAQC circuit. Simulations with MPS will only be efficient in systems satisfying an area law entanglement growth~\cite{SCHOLLWOCK201196}, which is far from the general case. As a naive alternative to overcome the requirements for $H_\text{S}$, one could also introduce random SQGs, albeit with no guarantees that the obtained $t_\text{A}$ is optimal \cite{MikelAlvaro2024}. 

While Eq.~\eqref{eq:arbitraryDAQCexponents} and Eq.~\eqref{eq:daqc_sand2} are always an exact identity, we see that Eq.~\eqref{eq:DAQCcircuit} will not longer be exact when working with arbitrary Hamiltonians, since the effective Hamiltonians do not necessarily commute with each other. Thus, solving this system of equations will not yield the ideal DAQC circuit, as there would be a Trotterization error. As a rule of thumb, we can employ the Lie formula to arbitrarily reduce the errors by dividing the circuit into different steps \cite{Adrian2020DAQC}, so that Eq.~\eqref{eq:DAQCcircuit} holds up to a controllable error. 

Importantly, from Eq.~\eqref{eq:daqc_sand2} it follows that the size of the system of equations needed to solve the problem grows with the total number of qubits. Even if we restrict ourselves to suboptimal protocols with polynomially large schedules, we would require large computational resources to obtain a valid DAQC circuit. This is one of the open problems that holds back the scaling of DAQC to systems consisting of a large number of qubits. Thus, there is still the question of whether we can obtain valid DAQC circuits using minimally costly algorithms.

\section{Constructive protocol for Hamiltonian simulation}\label{sec:result}

We introduce a new DAQC protocol that relies on an exact solution of Eq.~\eqref{eq:daqc_sand2} when the source Hamiltonian is of the form given by Eq.~\eqref{eq:zz_hamiltonians}. Our target is to obtain a valid DAQC circuit for simulating an arbitrary two-body Hamiltonian $H_\text{P}$ for a given time $T$ on a $N$ qubit system, minimizing computational resources. For that, we will first assume that $H_\text{S}$ is a ZZ Hamiltonian with compatible topology, i.e. ${h_\text{S}}_{ij}^{zz}\neq0$ if there is any coupling between qubits $i$ and $j$ in the problem Hamiltonian, ${h_\text{P}}_{ij}^{\mu\nu}\neq0$ for some $\mu,\nu$. 

While in previous attempts the usual approach was to define the digital blocks in terms of a finite set of SQGs~\cite{Bassler2023Synthesis,MikelAlvaro2024,bassler2025general}, here we allow for the implementation of arbitrary SQGs. Therefore, the unitary evolution corresponding to a digital block can be written as
\begin{eqnarray}\label{eq:lu_op}
U_k &=&U_{1k}\otimes...\otimes U_{ik}\otimes... \otimes U_{Nk},
\end{eqnarray}
where $U_{ik}$ represents an arbitrary single qubit gate for qubit $i$ in the $k$th DA block. In DAQC, we rely on the conjugation of the analog blocks with the digital blocks. In this case, we can obtain the effective Hamiltonian during a digital-analog block by analyzing its effect on a single qubit:
\begin{equation}\label{eq:Urotation}
    U_{ik} \sigma_i^z U_{ik}^\dagger =\gamma_{ik}^x\sigma_{i}^x+\gamma_{ik}^y\sigma_i^y+\gamma_{ik}^z\sigma_i^z,
\end{equation}
where the vector $\gamma_{ik}\coloneqq (\gamma_{ik}^x,\gamma_{ik}^y,\gamma_{ik}^z)$~\footnote{As a visual key, we will denote with the vector mark $\vec{\cdot}$ the vectors with size $3N$, in contrast to the vectors of size 3, which will be written without it.}, with $-1\leq\gamma_{ik}^\mu\leq1$ is normalized with respect to the 2-norm~\footnote{2-norm or euclidean norm. Being $v\in\mathbb{R}^n$, $\lVert v\rVert_2=(\sum_{i=1}^nv_i^2)^{1/2}$. As this is the only vector norm that we employ in this paper, we omit the sub-index.}
\begin{equation}\label{eq:normalization}
    \lVert\gamma_{ik}\rVert=\sqrt{({\gamma_{ik}^x})^2+({\gamma_{ik}^y})^2+({\gamma_{ik}^z})^2}=1 \hspace{5pt}\forall i,k, 
\end{equation}
and $\{\gamma_{ik}^x,\gamma_{ik}^y,\gamma_{ik}^z\}$ define the SQG (see Apx.~\ref{apx:SQG} for more details about the parametrization). With this we can write the effective source Hamiltonian for each block $k$, and the complete system of equations that follow from Eq.~\eqref{eq:daqc_sand2}. As in the usual DAQC fashion, we can take all constants to the right of the equation, grouping them in a single constant. For convenience, we choose to use the matrix notation 
\begin{equation}\label{eq:termbytermDAQC}
    \sum_k t_k\gamma_{ik}^\mu\gamma_{jk}^\nu=T{h_\text{P}}_{ij}^{\mu\nu}/{h_\text{S}}_{ij}^{zz}=\mathbf{B}_{ij}^{\mu\nu}, \forall i\neq j,\mu,\nu,
\end{equation}
where in general, $\mathbf{B}$ is a $3N\times3N$ real matrix.  This matrix can be divided into a matrix of $N\times N$ blocks, with each block being a submatrix of size $3\times3$. The submatrices are associated with the directional couplings between each pair of qubits. Notice that the values of the diagonal submatrices of $\mathbf{B}$ are not fixed, as they are not associated with any physical coupling. Notice also that the matrix $\mathbf{B}$ is symmetric by construction, even in the indeterminate diagonal blocks. 

Attempting to solve this equation using classical optimizers is a hard task, specially taking into account that we are facing an optimization problem with $9NK$ variables, where $K$ is the number of digital-analog blocks. In general we do not have a fixed value of $K$, but for arbitrary problems without symmetries we need at least as many variables as degrees of freedom to solve the equation, requiring $K\geq N(N-1)/2$. However, the constraint from Eq.~\eqref{eq:normalization} makes it difficult to navigate through the optimization landscape. Convergence of an heuristic approach might not be guaranteed and the computational cost to get a solution might be too high if convergence is not fast enough. Thus, we seek an analytical solution which requires minimal computational resources.

Let us start by noting that $\mathbf{B}$ is a Hermitian matrix, and thus we can obtain its eigendecomposition,
\begin{equation}
    \mathbf{B}=\mathbf{U}^\dagger \mathbf{\lambda}\mathbf{U},
\end{equation}
with $\mathbf{\lambda}$ a real diagonal matrix, and $\mathbf{U}$ a real unitary matrix with normalized columns $\vec{v}_l$. Notice that the form of the eigendecomposition of $\mathbf{B}$ is very similar to our problem in Eq.~\eqref{eq:termbytermDAQC}, albeit with some important differences. 

The first one, is that we need positive times for DAQC, so the eigenvalues $\lambda_k$ should all be positive. If we fix all indeterminate terms to be zero, the matrix $\mathbf{B}$ will be indefinite and will have a negative minimum eigenvalue $\tilde{\lambda}_\text{min}<0$. We can make $\mathbf{B}$ positive-semidefinite if we fix the diagonal to $-\tilde{\lambda}_\text{min}$, fixing the off-diagonal indeterminate elements to be zero. 

The second difference is complicated to solve. Usually, eigenvectors are given such that they are normalized, $\lVert \vec{v}_k\lVert=1$. However, we ask our vectors $\vec{\gamma}$ to fulfill a different normalization condition given by Eq.~\eqref{eq:normalization}. In general, there is no re-scaling of $\vec{v}$ that will satisfy this condition, as this would require each 3-element block of the vector to have a uniform norm, $\lVert v_i\rVert =\lVert v_j\rVert\ \forall i\neq j$, which is very unlikely to happen.

\begin{result}\label{result}
    For an arbitrary two-body Hamiltonian $H_\text{P}$ and a compatible ZZ Hamiltonian $H_\text{S}$, there exists a decomposition of $H_\text{P}$ which employs a quadratic number of transformations of $H_\text{S}$ with local unitary gates $U_q$,
    \begin{equation}
        TH_\text{P}=\sum_{q=1}^{\mathcal{O}(N^2)} t_qU_q H_\text{S} U_q^\dagger,\ t_q>0.
    \end{equation}
    Calculating this transformation requires only polynomial computational resources.
\end{result}

In other words, there is a decomposition of the matrix $\mathbf{B}$ with a polynomial number of steps and which employs a construction that only requires SQGs. The main idea behind this construction is to employ a divide and conquer strategy. Instead of focusing on the whole equation, we will search a solution for each of its eigenvectors. This way, we obtain a decomposition of the form
\begin{equation}\label{eq:fulldecomposition}
    \mathbf{B}=\sum_{k=1}^{3N}\lambda_k\vec{v}_k\vec{v}_k^\dagger=\sum_{k=1}^{3N} t_k\sum_{\ell=1}^{2N}\left(\vec{\gamma}_{+k}^{(\ell)}\vec{\gamma}_{+k}^{(\ell)\dagger}+\vec{\gamma}_{-k}^{(\ell)}\vec{\gamma}_{-k}^{(\ell)\dagger}\right),
\end{equation}
with $t_k=\lambda_k\max_i\lVert v_{ik}\rVert^2/(4N)$, $\vec{\gamma}_{\pm k}^{(\ell)}$ having components
\begin{equation}
    \gamma_{\pm ik}^{(\ell)}=\frac{v_{ik}\pm\varepsilon_{ik}^{(\ell)}}{\sqrt{\lVert v_{ik}\rVert^2+\lVert\varepsilon_{ik}^{(\ell)}\rVert^2}},
\end{equation}
and
\begin{equation}\label{eq:varepsilonparam}
    \varepsilon_{ik}^{(\ell)}=\cos\theta_{ik}^{(\ell)}\eta_{ik}+\sin\theta_{ik}^{(\ell)}\xi_{ik},
\end{equation}
with $v_{ik}\perp \eta_{ik}\perp \xi_{ik}$, $\lVert \eta_{ik}\rVert^2=\lVert \xi_{ik}\rVert^2=\max_i\lVert v_{ik}\rVert^2-\lVert v_{ik}\rVert^2$ and the angles given by 
\begin{equation}\label{eq:theta}
    \theta_{ik}^{(\ell)}=\frac{\pi(i-1)(\ell-1)}{N}\hspace{6pt}\forall k.
\end{equation}
Here the selection of the $\varepsilon_{ik}^{(\ell)}$ elements is not unique, but this particular choice allow us to provide a decomposition with $2N$ steps per eigenvalue. Indeed, $\eta_{ik}$ and $\xi_{ik}$ can take any particular value as long as they fulfill the conditions, and so are left as free parameters. Note here that the complexity of this construction is polynomial in time, as the computational cost is $\mathcal{O}(N^3)$, corresponding to the cost of the eigendecomposition of the problem matrix $\mathbf{B}$. A constructive proof is given in Apx.~\ref{apx:proof}.

The decomposition we have shown can be straightforwardly employed to generate a DAQC protocol for simulating $H_\text{P}$. Since we already have the analog block times, all we need to do is to extract information about the SQGs from the $\vec{\gamma}$ vectors. This depends on the particular choice parametrizing arbitrary SQGs. In Apx.~\ref{apx:SQG}, we show how to extract the angles of rotation for one particular SQG parametrization. With this, the DAQC circuit is:
\begin{equation}\label{eq:DAQCdecomposition}
    e^{-iTH_\text{P}} \approx\prod_{q} U_{q} e^{-it_qH_\text{S}}U_{q}^\dagger=\prod_{q} e^{-it_qH_\text{S}^{(q)}},
\end{equation}
with the index $q$ running over the indices employed in Eq.~\eqref{eq:fulldecomposition}. The derivation of the DAQC circuit is shown in Apx.~\ref{apx:DAQCcircuit}.

It is important to note that the target evolution operator and the one obtained from the DAQC circuit are not equal. Hence, we have a Trotter error comming from the decomposition of $H_\text{P}$ into non-commuting terms. As it is usual in such decompositions and DAQC, one can always use the Lie formula to divide the evolution into $n_\text{T}$ steps to reduce Trotterization errors. However, this might not be technically viable, as we would increase the number of digital-analog blocks by $n_\text{T}$ while also reducing the time of the analog block times by $t_k/n_\text{T}$.

All in all, the DAQC circuit is composed of a maximum of $12N^2$ digital-analog blocks, which is in the same order of magnitude as the ones required in previous protocols~\cite{Bassler2023Synthesis,MikelAlvaro2024,bassler2025general}. Also, the expression of the analog block times allows us to easily bound the total analog block time in terms of the eigenvalues of $\mathbf{B}$, 
\begin{equation}\label{eq:protocolupperbound}
t_\text{A}\leq\sum_k\lambda_k=\text{tr}(\mathbf{B})=3N\lvert\tilde{\lambda}_\text{min}\rvert.
\end{equation}
Additionally, this protocol gives us an immediate method for discarding negligible terms~\cite{Paula2024, MikelAlvaro2024}. In particular, we can discard the simulation corresponding to the $k$-th eigenvector if the associated analog time $t_k$ is below certain threshold.

\section{Numerical calculations}\label{sec:numerics}

Here we show results on the calculations performed to verify the scalability of our approach. For that, first we have to generate valid problems. In our case, we have chosen to generate the $\mathbf{B}$ directly, without connecting it to any particular problem nor source Hamiltonians. For our tests, we have generated these matrices with a random uniform distribution on the range $-1\leq\mathbf{B}_{3i+\mu,3j+\nu}\leq1$. This ensures that the generated matrices represent systems without symmetries or other exploitable features, thus corresponding to the worst possible case for DAQC compilation. We further normalize our matrices such that the maximum magnitude of any element is 1, i.e. $\max\lvert \mathbf{B}_{3i+\mu,3j+\nu}\rvert=1$. With these problem matrices, we have computed the total analog block times $t_\text{A}$ obtained using the construction outlined in Result~\ref{result}. 

\begin{figure}[!t]
    \centering
    \includegraphics[width=\linewidth]{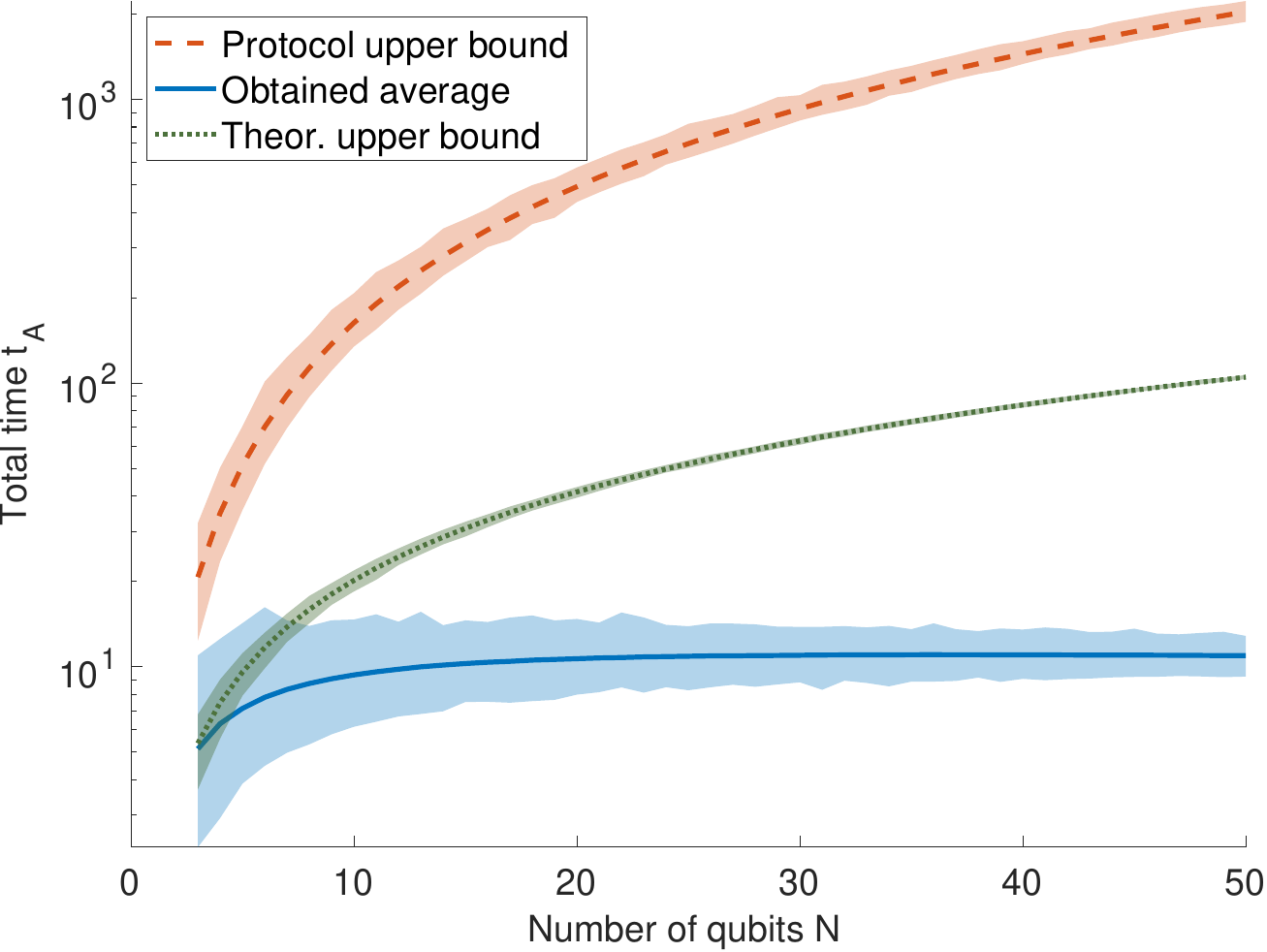}
    \caption{$\log t_\text{A}$ for random problems of size $N$. The problem matrices ${\textbf{B}}$ are generated with uniform values in the range $[-1,1]$ and then normalized with the elementwise infinite norm. The solid line shows the average obtained over $10^4$ runs. The dashed line shows the upper bound calculated as $t_\text{A}\leq 3N\lvert\tilde{\lambda}_\text{min}\rvert$. The dotted line represents the theoretical upper bound for the optimal DAQC protocol with the Pauli gates. The colored regions represent the area between the maximum and minimum values obtained.}
    \label{fig:taVSn}
\end{figure}

In Fig.~\ref{fig:taVSn}, we represent $t_\text{A}$ as a function of the total number of qubits $N$, up to systems of size $N=50$. For each value of $N$, we have generated $10^4$ random $\mathbf{B}$ matrices. In the figure, we show the obtained total time, and the upper bound given by Eq.\eqref{eq:protocolupperbound}. In order to compare this protocol against previous ones, we also represent the obtained upper bound of $t_\text{A}$ for an optimal DAQC protocol as provided in Ref.~\cite{Mikel2025bounds}, in this case  $t_\text{A}\leq\sqrt{3}(\sum_{i<j}\mathbf{B}_{i,j}^2)^{1/2}$ with the elements in the diagonal blocks equal to 0.

Fig.~\ref{fig:taVSn} shows that using this normalization of ${\textbf{B}}$, $t_\text{A}$ scales nearly constant with system size. If we compare both upper bounds, we see that the upper bound for this protocol is significantly larger than the theoretical upper bound for optimal DAQC protocols. This is expected, as the protocol provided in this work is suboptimal and penalizes the optimal upper bound by some factor. Nevertheless, this factor seems to increase slower than linearly with the number of qubits. Additionally, we observe that the obtained $t_\text{A}$ grows linearly with the maximum magnitude of the ratio between the problem and the source Hamiltonians, $t_\text{A}\sim T\max\lvert \mathbf{B}_{3i+\mu,3j+\nu}\rvert= T\max\lvert {h_\text{P}}_{ij}^{\mu\nu}\oslash {h_\text{S}}_{ij}^{\mu\nu}\rvert$.
In this case, as this value is chosen to be constant, Fig.\ref{fig:taVSn} flattens as $N$ grows.

Although we have not studied extreme cases in which the upper bound for $t_\text{A}$ saturates, and our numerical calculations focused on a specific distribution of problems, we expect usual DAQC compilation tasks to involve problem matrices $\text{B}$ following a distribution of their elements similar to the one employed here. 

\section{Conclusions}\label{sec:conclusion}
The main result of this paper provides a systematic way for obtaining decompositions of arbitrary two-body Hamiltonians using a set of local unitary transformations applied to a ZZ Hamiltonian. The novelty of this approach is that it allows to obtain a DAQC protocol employing minimal computational resources, providing a recipe-based method to design DAQC circuits that simulate unitary evolution under arbitrary two-body Hamiltonians. In contrast to previous approaches where solving a hard optimization problem at the preprocessing stage is necessary, the protocol presented in this work eliminates this requirement by making use of exact expressions. Indeed, we have shown that the original problem can be brought into a form similar to eigendecomposition of a $3N\times 3N$ positive semidefinite matrix, with $N$ the number of qubits, thus yielding a runtime polynomial with system size, $\mathcal{O}(N^3)$. Additionally, the proposed protocol yields DAQC circuits composed of at most $12N^2$ digital analog blocks, which is in the same order of magnitude as previous protocols that required up to $9N(N-1)/2$ blocks. Even though the presented protocol is suboptimal in terms of total analog block time $t_\text{A}=\sum_q t_q$, the numerical calculations reported here show that this value does not grow with $N$ for problems in which the ratio between the problem and the source Hamiltonian couplings is bounded, suggesting a relation $t_\text{A}\sim T\max\lvert {h_\text{P}}_{ij}^{\mu\nu}\oslash {h_\text{S}}_{ij}^{\mu\nu}\rvert$. These results contribute to a favourable scaling of DAQC protocols for systems consisting of a large number of qubits, in this case, by lowering the classical computational resources needed for compiling such circuits.

Although the protocol presented here relies on source Hamiltonians having only ZZ terms, this can be easily extended to Hamiltonians with symmetric terms (i.e. Hamiltonians with XX or YY terms). For generic source Hamiltonians, one could first obtain an effective ZZ Hamiltonian from the protocol by interchanging the roles of $H_\text{S}$ and $H_\text{P}$, and then apply the protocol again to simulate the problem of interest. However, this would introduce a large overhead in the number of required blocks~\footnote{As we would need to obtain an effective ZZ Hamiltonian for each block, we would have a total number of blocks in the order of $\mathcal{O}(N^4)$.}. Extension of the protocol to efficiently employ arbitrary source Hamiltonians as those in Eq.~\eqref{eq:daqc_sand2} is still an open question. 

\section*{Acknowledgements}
We acknowledge discussions with Jochen Szangolies during preparation of the manuscript. This project was made possible by the DLR Quantum Computing Initiative and the Federal Ministry for Economic Affairs and Climate Action of Germany; \url{qci.dlr.de/projects/IQDA}. MGdA acknowledges support from the HORIZONCL4- 2022-QUANTUM01-SGA project 101113946 OpenSuperQPlus100 of the EU Flagship on Quantum Technologies, the project Grant No. PID2024-156808NB-I00 and Spanish Ramón y Cajal Grant No. RYC-2020-030503-I funded by MICIU/AEI/10.13039/501100011033 and by “ERDF A way of making Europe” and “ERDF Invest in your Future”, and the  Basque Government through Grant No. IT1470-22 and Elkartek project KUBIBIT-kuantikaren berrikuntzarako ibilbide teknologikoak (ELKARTEK25/79).

\appendix
\section{SQG parametrization}\label{apx:SQG}

Several parametrizations can be chosen for the SQGs. Here, we define an arbitrary single qubit rotation as
\begin{equation}
R(\theta,\hat{n})=e^{-i\frac{\theta}{2}\left(n_{x}\sigma^x+n_{y}\sigma^y+n_{z}\sigma^z\right)},
\end{equation}
where $\hat{n}=(n_{x},n_{y},n_{z})$ is a unit vector. This operator can be employed to construct each of the local unitaries in Eq.~\eqref{eq:lu_op}; we have omitted the qubit index to ease the notation. Applying this arbitrary rotation to Eq.~\eqref{eq:Urotation}, we obtain the following identities
\begin{align}
&\gamma^x = n_{x}n_{z}(1-\cos(\theta))+n_{y}\sin(\theta),\nonumber\\
&\gamma^y = n_{y}n_{z}(1-\cos(\theta))-n_{x}\sin(\theta),\nonumber\\
&\gamma^z = n_{z}^2(1-\cos(\theta))+\cos(\theta)
\end{align}
Note that due to the normalization condition $\sqrt{n_{x}^2+n_{y}^2+n_{z}^2}=1$, there are three degrees of freedom for a single arbitrary rotation. Further note that here we can choose $n_z=0$, simplifying the identities.

\section{Proof of main result}\label{apx:proof}

In this appendix we give a proof for Result \ref{result}. For that, we will construct step by step a decomposition for any problem $\mathbf{B}$ as given in Eq.~\eqref{eq:termbytermDAQC}.

As we have seen in the main text, we can always write the problem as a positive semi-definite matrix by giving a value to the indeterminate terms in the diagonal blocks. This way, our goal is to find a decomposition of the form
\begin{equation}\label{eq:originalproblem}
    \mathbf{B}=\sum_{k=1}^{3N}\lambda_k\vec{v}_k\vec{v}_k^\dagger=\sum_{q=1}^Qt_q\vec{\gamma}_q\vec{\gamma}_q^\dagger,
\end{equation}
with $\lambda_k\geq0$ the eigenvalues corresponding to the eigenvector $\vec{v}_k$, $t_q>0$ and $\vec{\gamma}_q$ subjected to the normalization condition given in Eq.~\eqref{eq:normalization}.In the presence of degenerate eigenvalues, the corresponding eigenvectors $v_k$ should be chosen such that their corresponding $\lVert v_i\rVert$ is non-zero such that the normalization condition can be met.

Instead of focusing on the whole equation, we will search a solution for each of the eigenvector projectors,
\begin{equation}\label{eq:singleeigenvector}
    \lambda \vec{v}\vec{v}^\dagger=\sum_{\ell=1}^L t^{(\ell)}\vec{\gamma}^{(\ell)}\vec{\gamma}^{(\ell)\dagger},
\end{equation}
where we have dropped the index $k$ to ease the notation.

As an initial guess, we consider $\vec{\gamma}^{(\ell)}$ vectors consisting of different blocks given by the eigenvector components plus some perpendicular perturbation $\varepsilon_i^{(\ell)}\perp v_i$ and which fulfill Eq.~\eqref{eq:normalization},
\begin{equation} 
    \gamma_i^{(\ell)} = \frac{v_i+\varepsilon_i^{(\ell)}}{\lVert v_i+\varepsilon_i^{(\ell)}\rVert}=\frac{v_i+\varepsilon_i^{(\ell)}}{\sqrt{\lVert v_i\rVert^2+\lVert\varepsilon_i^{(\ell)}\rVert^2}}.
\end{equation}
As it is not possible to solve Eq.~\eqref{eq:singleeigenvector} with a single step, we begin with two steps. This way, for each $i\neq j$ we have
\begin{eqnarray}
    \lambda v_iv_j^\dagger = t^{(1)}\gamma_i^{(1)}\gamma_j^{(1)\dagger}+t^{(2)}\gamma_i^{(2)}\gamma_j^{(2)\dagger}.
\end{eqnarray}

In order to make the equation simpler, we choose to cancel the crossed terms resulting from $v_i\varepsilon_j^{(\ell)}$ and $\varepsilon_i^{(\ell)}v_j$ by selecting the vectors such that $\vec{\varepsilon}^{(1)}=-\vec{\varepsilon}^{(2)}=\vec{\varepsilon}$. This fixes the election of the times to cancel the crossed terms, $t^{(1)}=t^{(2)}=t$. Let us further impose that $\lVert v_i\rVert^2+\lVert\varepsilon_i\rVert^2=\lVert v_j\rVert^2+\lVert\varepsilon_j\rVert^2\ \forall i\neq j$. This choices leaves us with
\begin{equation}
    \lambda v_iv_j^\dagger=\frac{2t}{\lVert v_i\rVert^2+\lVert\varepsilon_{i}\rVert^2}\left(v_iv_j^\dagger+\varepsilon_i\varepsilon_j^\dagger\right).
\end{equation}
This equation suggests fixing the analog block times and the norms such that
\begin{equation}\label{eq:fix_norms}
    \lVert v_i\rVert^2+\lVert\varepsilon_{i}\rVert^2 = \frac{2t}{\lambda}\hspace{5pt}\forall i.
\end{equation}
From this equation for two steps, we can extract a lower bound on the analog block times, $t\geq \lambda \max_i\lVert v_i\rVert^2/2$. As we would like to minimize the total circuit time, we can set the value of $t$ to this limiting value. This fixes the norms of all $\varepsilon_i$ vectors to $\lVert \varepsilon_i\rVert^2=\max_i\lVert v_i\rVert^2-\lVert v_i\rVert^2$.

Up to this point, we have obtained an approximation of $\lambda \vec{v}\vec{v}^\dagger$ with an error of $\lambda\vec{\varepsilon}\vec{\varepsilon}^\dagger$. This hints to employ more steps in order to obtain an exact expansion of our target. In doing so, we will construct pairs of blocks in which we select the $\vec{\varepsilon}^{(\ell)}$ vectors in the way discussed previously for $L$ different contributions. The decomposition takes the form:
\begin{equation}\label{eq:ellsteps}
    \lambda v_iv_j^\dagger=\sum_{\ell=1}^L\frac{2t^{(\ell)}}{\lVert v_i\rVert^2+\lVert\varepsilon_{i}^{(\ell)}\rVert^2}\left(v_iv_j^\dagger+\varepsilon_i^{(\ell)}\varepsilon_j^{(\ell)\dagger}\right),
\end{equation}
with $\varepsilon_i^{(\ell)}\perp v_i$. Now, we take an extra choice by setting all times to be equal, $t^{(\ell)}=t$. This way, we fix the norms of the vectors for all steps, $\lVert\varepsilon_i^{(\ell)}\rVert=\lVert\varepsilon_i\rVert$. For a total of $L$ steps, we take $t=\lambda \max_i\lVert v_i\rVert^2/(2L)$ to fix the equality on the $v_iv_j^\dagger$ terms.

The last step to fulfill the identity in Eq.~\eqref{eq:ellsteps} is to fix the remaining free parameters. For that, we expand the rightmost term of Eq.~\eqref{eq:ellsteps} to the the following system of equations
\begin{equation}\label{eq:finalsystemofequations}
    \sum_{\ell=1}^L\varepsilon_i^{(\ell)}\varepsilon_j^{(\ell)\dagger}=0_{3,3}\ \forall i\neq j.
\end{equation}
Since we have previously fixed the norm of these vectors and the plane they live in, we can parameterize them with a single parameter:
\begin{equation}\label{eq:varepsilonparam}
    \varepsilon_i^{(\ell)}=\cos\theta_i^{(\ell)}\eta_i+\sin\theta_i^{(\ell)}\xi_i,
\end{equation}
such that $v_i\perp\eta_i\perp\xi_i$, $\lVert\varepsilon_i\rVert=\lVert\eta_i\rVert=\lVert\xi_i\rVert$, and $\theta_i^{(\ell)}\in[0,2\pi)$. Note here that these conditions can be satisfied even in the cases in which $v_i=0$, as in this case we can consider that any vector is orthogonal to it, $\vec{0}\perp\vec{a}\ \forall \vec{a}$. Here, we have freedom for selecting any auxiliary vectors $\eta_i$ and $\xi_i$ at qubit $i$ that fulfill these conditions. An easy analytical solution to this system of equations with a number of steps $L=2N$ \footnote{As a fun fact, if we wanted to solve this equation with the optimum amount of steps, there is another analytical solution which employs $N$ steps if $N$ is a power of 2. In this case, we can associate the matrix $\vec{\varepsilon}\vec{\varepsilon}^\dagger$ with the $n\times n$ Hadamard matrix, as its definition coincides with Eq.~\eqref{eq:finalsystemofequations} for all off-diagonal $3\times3$ blocks.} is found by choosing:
\begin{equation}\label{eq:theta}
    \theta_i^{(\ell)}=\frac{\pi(i-1)(\ell-1)}{N}.
\end{equation}
This way, we can build the decomposition of the matrix $\mathbf{B}$ shown in Eq.~\eqref{eq:fulldecomposition}.

\section{From the decomposition of $\mathbf{B}$ to the DAQC circuit}\label{apx:DAQCcircuit}

In this Appendix we show how to construct the DAQC circuit that simulates the evolution under $H_\text{P}$ by employing the decomposition from Eq.~\eqref{eq:fulldecomposition}.

We start from the definition of the matrix $\textbf{B}$ given in Eq.~\eqref{eq:termbytermDAQC}, and we isolate the couplings from the problem Hamiltonian
\begin{equation}
    T{h_\text{P}}_{ij}^{\mu\nu}={h_\text{S}}_{ij}^{zz}\mathbf{B}_{ij}^{\mu\nu},\ \forall i\neq j, \mu,\nu.
\end{equation}
Now, we can substitute the expression from Eq.~\eqref{eq:fulldecomposition},
\begin{equation}
\begin{split}
    T{h_\text{P}}_{ij}^{\mu\nu}&={h_\text{S}}_{ij}^{zz}\sum_{k=1}^{3N} t_k\sum_{\ell=1}^{2N}\left(\vec{\gamma}_{+ik}^{(\ell)\mu}\vec{\gamma}_{+jk}^{(\ell)\nu}+\vec{\gamma}_{-ik}^{(\ell)\mu}\vec{\gamma}_{-jk}^{(\ell)\nu}\right)\\
    &=\sum_q t_q{h_\text{S}}_{ij}^{\mu\nu(q)}.
\end{split}
\end{equation}
where ${h_\text{S}}_{ij}^{\mu\nu (q)}$ is the effective coupling of the source Hamiltonian at the $q$th step. Adding the corresponding Pauli terms and summing over all indices, we can write the equation corresponding to the decomposition of the problem Hamiltonian as in Eq.~\eqref{eq:daqc_sand2},
\begin{align}    
    TH_\text{P}&=T\sum_{i<j}\sum_{\mu,\nu}{h_\text{P}}_{ij}^{\mu\nu}\sigma_i^\mu\sigma_j^\nu=\sum_{k=1}t_k H_\text{P}^{(k)}\nonumber\\
    &=\sum_{i<j}\sum_{\mu,\nu}{h_\text{S}}_{ij}^{zz}\sum_{\eta=\{+,-\}}\sum_{k=1}^{3N} t_k\sum_{\ell=1}^{2N}\vec{\gamma}_{\eta ik}^{(\ell)\mu}\vec{\gamma}_{\eta jk}^{(\ell)\nu}\sigma_i^\mu\sigma_j^\nu\nonumber\\
    &=\sum_q t_qU_q H_\text{S}U_q^\dagger=\sum_q t_qH_\text{S}^{(q)}.
\end{align}

The last step is employ Eq.~\eqref{eq:Urotation} to obtain the local unitary rotations $U_q$ giving the effective Hamiltonian. A mapping like the one described in Apx.\ref{apx:SQG} can be used to obtain the parameters of the single qubit gates generating each effective Hamiltonian. With this, we arrive at the expression shown in Eq.~\eqref{eq:DAQCdecomposition}.

We stress that the gauge invariance employed here to rewrite the matrix $\mathbf{B}$ as a positive semidefinite matrix has no effect in the overall Hamiltonian decomposition. This is because indeterminate values of $\mathbf{B}$ correspond to two equal qubit indices. Since these do not correspond to two-body terms, changing these indeterminate values does not alter the description of the original problem.

\bibliography{main.bib}

@Article{Lloyd1996,
  author   = {Seth Lloyd },
  journal  = {Science},
  title    = {Universal Quantum Simulators},
  year     = {1996},
  number   = {5278},
  pages    = {1073-1078},
  volume   = {273},
  doi      = {10.1126/science.273.5278.1073},
  url      = {https://www.science.org/doi/abs/10.1126/science.273.5278.1073},
}

@Article{Buluta2009,
  author   = {Iulia Buluta and Franco Nori },
  journal  = {Science},
  title    = {Quantum Simulators},
  year     = {2009},
  number   = {5949},
  pages    = {108-111},
  volume   = {326},
  doi      = {10.1126/science.1177838},
  url      = {https://www.science.org/doi/abs/10.1126/science.1177838},
}

@Article{Cirac2012,
  author   = {Cirac, J. Ignacio and Zoller, Peter},
  journal  = {Nature Physics},
  title    = {Goals and opportunities in quantum simulation},
  year     = {2012},
  issn     = {1745-2481},
  number   = {4},
  pages    = {264--266},
  volume   = {8},
  doi      = {10.1038/nphys2275},
  refid    = {Cirac2012},
  url      = {https://doi.org/10.1038/nphys2275},
}

@article{Monroe2021,
  title = {Programmable quantum simulations of spin systems with trapped ions},
  author = {Monroe, C. and Campbell, W. C. and Duan, L.-M. and Gong, Z.-X. and Gorshkov, A. V. and Hess, P. W. and Islam, R. and Kim, K. and Linke, N. M. and Pagano, G. and Richerme, P. and Senko, C. and Yao, N. Y.},
  journal = {Rev. Mod. Phys.},
  volume = {93},
  issue = {2},
  pages = {025001},
  numpages = {57},
  year = {2021},
  month = {Apr},
  publisher = {American Physical Society},
  doi = {10.1103/RevModPhys.93.025001},
  url = {https://link.aps.org/doi/10.1103/RevModPhys.93.025001}
}

@Article{Blatt2012,
  author   = {Blatt, R. and Roos, C. F.},
  journal  = {Nature Physics},
  title    = {Quantum simulations with trapped ions},
  year     = {2012},
  issn     = {1745-2481},
  number   = {4},
  pages    = {277--284},
  volume   = {8},
  doi      = {10.1038/nphys2252},
  refid    = {Blatt2012},
  url      = {https://doi.org/10.1038/nphys2252},
}

@Article{Barreiro2011,
  author   = {Barreiro, Julio T. and Müller, Markus and Schindler, Philipp and Nigg, Daniel and Monz, Thomas and Chwalla, Michael and Hennrich, Markus and Roos, Christian F. and Zoller, Peter and Blatt, Rainer},
  journal  = {Nature},
  title    = {An open-system quantum simulator with trapped ions},
  year     = {2011},
  issn     = {1476-4687},
  number   = {7335},
  pages    = {486--491},
  volume   = {470},
  doi      = {10.1038/nature09801},
  refid    = {Barreiro2011},
  url      = {https://doi.org/10.1038/nature09801},
}

@Article{Bloch2012,
  author   = {Bloch, Immanuel and Dalibard, Jean and Nascimbène, Sylvain},
  journal  = {Nature Physics},
  title    = {Quantum simulations with ultracold quantum gases},
  year     = {2012},
  issn     = {1745-2481},
  number   = {4},
  pages    = {267--276},
  volume   = {8},
  doi      = {10.1038/nphys2259},
  refid    = {Bloch2012},
  url      = {https://doi.org/10.1038/nphys2259},
}

@Article{Gross2017,
  author   = {Christian Gross and Immanuel Bloch },
  journal  = {Science},
  title    = {Quantum simulations with ultracold atoms in optical lattices},
  year     = {2017},
  number   = {6355},
  pages    = {995-1001},
  volume   = {357},
  doi      = {10.1126/science.aal3837},
  url      = {https://www.science.org/doi/abs/10.1126/science.aal3837},
}

@Article{Bernien2017,
  author   = {Bernien, Hannes and Schwartz, Sylvain and Keesling, Alexander and Levine, Harry and Omran, Ahmed and Pichler, Hannes and Choi, Soonwon and Zibrov, Alexander S. and Endres, Manuel and Greiner, Markus and Vuletić, Vladan and Lukin, Mikhail D.},
  journal  = {Nature},
  title    = {Probing many-body dynamics on a 51-atom quantum simulator},
  year     = {2017},
  issn     = {1476-4687},
  number   = {7682},
  pages    = {579--584},
  volume   = {551},
  doi      = {10.1038/nature24622},
  refid    = {Bernien2017},
  url      = {https://doi.org/10.1038/nature24622},
}

@Article{Ebadi2021,
  author   = {Ebadi, Sepehr and Wang, Tout T. and Levine, Harry and Keesling, Alexander and Semeghini, Giulia and Omran, Ahmed and Bluvstein, Dolev and Samajdar, Rhine and Pichler, Hannes and Ho, Wen Wei and Choi, Soonwon and Sachdev, Subir and Greiner, Markus and Vuletić, Vladan and Lukin, Mikhail D.},
  journal  = {Nature},
  title    = {Quantum phases of matter on a 256-atom programmable quantum simulator},
  year     = {2021},
  issn     = {1476-4687},
  number   = {7866},
  pages    = {227--232},
  volume   = {595},
  doi      = {10.1038/s41586-021-03582-4},
  refid    = {Ebadi2021},
  url      = {https://doi.org/10.1038/s41586-021-03582-4},
}

@article{Dodd2002UnivQC,
title = {Universal quantum computation and simulation using any entangling Hamiltonian and local unitaries},
author = {Dodd, Jennifer L. and Nielsen, Michael A. and Bremner, Michael J. and Thew, Robert T.},
journal = {Phys. Rev. A},
volume = {65},
issue = {4},
pages = {040301(R)},
numpages = {4},
year = {2002},
publisher = {American Physical Society},
doi = {10.1103/PhysRevA.65.040301},
url = {https://link.aps.org/doi/10.1103/PhysRevA.65.040301}
}

@article{Adrian2020DAQC,
title={Digital-analog quantum computation},
volume={101},
url={http://dx.doi.org/10.1103/PhysRevA.101.022305},
DOI={10.1103/physreva.101.022305},
number={2},
pages={022305},
year={2020},
journal={Phys. Rev. A},
publisher={American Physical Society (APS)},
author={Parra-Rodriguez, Adrian and Lougovski, Pavel and Lamata, Lucas and Solano, Enrique and Sanz, Mikel}
}

@article{MikelAlvaro2024,
  title = {Digital-analog quantum computation with arbitrary two-body Hamiltonians},
  author = {Garcia-de-Andoin, Mikel and Saiz, \'Alvaro and P\'erez-Fern\'andez, Pedro and Lamata, Lucas and Oregi, Izaskun and Sanz, Mikel},
  journal = {Phys. Rev. Res.},
  volume = {6},
  issue = {1},
  pages = {013280},
  numpages = {14},
  year = {2024},
  month = {Mar},
  publisher = {American Physical Society},
  doi = {10.1103/PhysRevResearch.6.013280},
  url = {https://link.aps.org/doi/10.1103/PhysRevResearch.6.013280}
}

@article{Bassler2023Synthesis,
   title={Synthesis of and compilation with time-optimal multi-qubit gates},
   volume={7},
   ISSN={2521-327X},
   url={http://dx.doi.org/10.22331/q-2023-04-20-984},
   DOI={10.22331/q-2023-04-20-984},
   journal={Quantum},
   publisher={Verein zur Forderung des Open Access Publizierens in den Quantenwissenschaften},
   author={Baßler, Pascal and Zipper, Matthias and Cedzich, Christopher and Heinrich, Markus and Huber, Patrick H. and Johanning, Michael and Kliesch, Martin},
   year={2023},
   month=apr, 
   pages={984} 
}

@misc{YanJunchi2024review,
      title={Quantum Circuit Synthesis and Compilation Optimization: Overview and Prospects}, 
      author={Yan Ge and Wu Wenjie and Chen Yuheng and Pan Kaisen and Lu Xudong and Zhou Zixiang and Wang Yuhan and Wang Ruocheng and Yan Junchi},
      year={2024},
      eprint={2407.00736},
      archivePrefix={arXiv},
      primaryClass={quant-ph},
      url={https://arxiv.org/abs/2407.00736}, 
}

@article{Lloyd1995Univesal,
  title = {Almost Any Quantum Logic Gate is Universal},
  author = {Lloyd, Seth},
  journal = {Phys. Rev. Lett.},
  volume = {75},
  issue = {2},
  pages = {346--349},
  numpages = {0},
  year = {1995},
  month = {Jul},
  publisher = {American Physical Society},
  doi = {10.1103/PhysRevLett.75.346},
  url = {https://link.aps.org/doi/10.1103/PhysRevLett.75.346}
}

@article{Barenco1995Elementary,
  title = {Elementary gates for quantum computation},
  author = {Barenco, Adriano and Bennett, Charles H. and Cleve, Richard and DiVincenzo, David P. and Margolus, Norman and Shor, Peter and Sleator, Tycho and Smolin, John A. and Weinfurter, Harald},
  journal = {Phys. Rev. A},
  volume = {52},
  issue = {5},
  pages = {3457--3467},
  numpages = {0},
  year = {1995},
  month = {Nov},
  publisher = {American Physical Society},
  doi = {10.1103/PhysRevA.52.3457},
  url = {https://link.aps.org/doi/10.1103/PhysRevA.52.3457}
}

@Article{Paula2024,
author={Garc{\'i}a-Molina, Paula and Martin, Ana and Garcia de Andoin, Mikel and Sanz, Mikel},
title={Mitigating noise in digital and digital--analog quantum computation},
journal={Communications Physics},
year={2024},
month={Oct},
day={06},
volume={7},
number={1},
pages={321},
issn={2399-3650},
doi={10.1038/s42005-024-01812-5},
url={https://doi.org/10.1038/s42005-024-01812-5}
}

@article{van2025optimal,
  title={Optimal compilation of parametrised quantum circuits},
  author={van de Wetering, John and Yeung, Richie and Laakkonen, Tuomas and Kissinger, Aleks},
  journal={Quantum},
  volume={9},
  pages={1828},
  year={2025},
  publisher={Verein zur F{\"o}rderung des Open Access Publizierens in den Quantenwissenschaften}
}

@Article{Haferkamp2022,
author={Haferkamp, Jonas and Faist, Philippe and Kothakonda, Naga B. T. and Eisert, Jens and Yunger Halpern, Nicole},
title={Linear growth of quantum circuit complexity},
journal={Nature Physics},
year={2022},
month={May},
day={01},
volume={18},
number={5},
pages={528-532},
issn={1745-2481},
doi={10.1038/s41567-022-01539-6},
url={https://doi.org/10.1038/s41567-022-01539-6}
}

@Article{Moro2021,
author={Moro, Lorenzo and Paris, Matteo G. A. and Restelli, Marcello and Prati, Enrico},
title={Quantum compiling by deep reinforcement learning},
journal={Communications Physics},
year={2021},
month={Aug},
day={06},
volume={4},
number={1},
pages={178},
issn={2399-3650},
doi={10.1038/s42005-021-00684-3},
url={https://doi.org/10.1038/s42005-021-00684-3}
}

@article{Nori2012,
   title={Two-qubit gate operations in superconducting circuits with strong coupling and weak anharmonicity},
   volume={14},
   ISSN={1367-2630},
   url={http://dx.doi.org/10.1088/1367-2630/14/7/073041},
   DOI={10.1088/1367-2630/14/7/073041},
   number={7},
   journal={New Journal of Physics},
   publisher={IOP Publishing},
   author={Lü, Xin-You and Ashhab, S and Cui, Wei and Wu, Rebing and Nori, Franco},
   year={2012},
   month=jul, 
   pages={073041}
}

@article{Joel2023,
  title = {Implementing two-qubit gates at the quantum speed limit},
  author = {Howard, Joel and Lidiak, Alexander and Jameson, Casey and Basyildiz, Bora and Clark, Kyle and Zhao, Tongyu and Bal, Mustafa and Long, Junling and Pappas, David P. and Singh, Meenakshi and Gong, Zhexuan},
  journal = {Phys. Rev. Res.},
  volume = {5},
  issue = {4},
  pages = {043194},
  numpages = {11},
  year = {2023},
  month = {Dec},
  publisher = {American Physical Society},
  doi = {10.1103/PhysRevResearch.5.043194},
  url = {https://link.aps.org/doi/10.1103/PhysRevResearch.5.043194}
}

@article{chen2025efficient,
  title={Efficient implementation of arbitrary two-qubit gates using unified control},
  author={Chen, Zhen and Liu, Weiyang and Ma, Yanjun and Sun, Weijie and Wang, Ruixia and Wang, He and Xu, Huikai and Xue, Guangming and Yan, Haisheng and Yang, Zhen and others},
  journal={Nature Physics},
  pages={1--8},
  year={2025},
  publisher={Nature Publishing Group UK London}
}

@article{Shapira2023,
  title = {Robust Two-Qubit Gates for Trapped Ions Using Spin-Dependent Squeezing},
  author = {Shapira, Yotam and Cohen, Sapir and Akerman, Nitzan and Stern, Ady and Ozeri, Roee},
  journal = {Phys. Rev. Lett.},
  volume = {130},
  issue = {3},
  pages = {030602},
  numpages = {7},
  year = {2023},
  month = {Jan},
  publisher = {American Physical Society},
  doi = {10.1103/PhysRevLett.130.030602},
  url = {https://link.aps.org/doi/10.1103/PhysRevLett.130.030602}
}

@article{Galicia2020,
  title = {Enhanced connectivity of quantum hardware with digital-analog control},
  author = {Galicia, Asier and Ramon, Borja and Solano, Enrique and Sanz, Mikel},
  journal = {Phys. Rev. Res.},
  volume = {2},
  issue = {3},
  pages = {033103},
  numpages = {11},
  year = {2020},
  month = {Jul},
  publisher = {American Physical Society},
  doi = {10.1103/PhysRevResearch.2.033103},
  url = {https://link.aps.org/doi/10.1103/PhysRevResearch.2.033103}
}

@article{Bassler2024timeoptimalmulti,
  doi = {10.22331/q-2024-03-13-1279},
  url = {https://doi.org/10.22331/q-2024-03-13-1279},
  title = {Time-optimal multi-qubit gates: {C}omplexity, efficient heuristic and gate-time bounds},
  author = {Ba{\ss{}}ler, Pascal and Heinrich, Markus and Kliesch, Martin},
  journal = {{Quantum}},
  issn = {2521-327X},
  publisher = {{Verein zur F{\"{o}}rderung des Open Access Publizierens in den Quantenwissenschaften}},
  volume = {8},
  pages = {1279},
  month = mar,
  year = {2024}
}

@article{Amy2014,
   title={Polynomial-Time T-Depth Optimization of Clifford+T Circuits Via Matroid Partitioning},
   volume={33},
   ISSN={1937-4151},
   url={http://dx.doi.org/10.1109/TCAD.2014.2341953},
   DOI={10.1109/tcad.2014.2341953},
   number={10},
   journal={IEEE Transactions on Computer-Aided Design of Integrated Circuits and Systems},
   publisher={Institute of Electrical and Electronics Engineers (IEEE)},
   author={Amy, Matthew and Maslov, Dmitri and Mosca, Michele},
   year={2014},
   month=oct, 
   pages={1476–1489} 
}

@Article{Ruiz2025,
author={Ruiz, Francisco J. R. and Laakkonen, Tuomas and Bausch, Johannes and Balog, Matej and Barekatain, Mohammadamin and Heras, Francisco J. H. and Novikov, Alexander and Fitzpatrick, Nathan and Romera-Paredes, Bernardino and van de Wetering, John and Fawzi, Alhussein and Meichanetzidis, Konstantinos and Kohli, Pushmeet},
title={Quantum circuit optimization with AlphaTensor},
journal={Nature Machine Intelligence},
year={2025},
month={Mar},
day={01},
volume={7},
number={3},
pages={374-385},
issn={2522-5839},
doi={10.1038/s42256-025-01001-1},
url={https://doi.org/10.1038/s42256-025-01001-1}
}

@article{SCHOLLWOCK201196,
title = {The density-matrix renormalization group in the age of matrix product states},
journal = {Annals of Physics},
volume = {326},
number = {1},
pages = {96-192},
year = {2011},
note = {January 2011 Special Issue},
issn = {0003-4916},
doi = {https://doi.org/10.1016/j.aop.2010.09.012},
url = {https://www.sciencedirect.com/science/article/pii/S0003491610001752},
author = {Ulrich Schollwöck}
}

@misc{yan2025,
      title={Quantum Circuit Synthesis and Compilation Optimization: Overview and Prospects}, 
      author={Ge Yan and Wenjie Wu and Yuheng Chen and Kaisen Pan and Xudong Lu and Zixiang Zhou and Yuhan Wang and Ruocheng Wang and Junchi Yan},
      year={2025},
      eprint={2407.00736},
      archivePrefix={arXiv},
      primaryClass={quant-ph},
      url={https://arxiv.org/abs/2407.00736}, 
}

@misc{kjelstrom2025exactquantumcircuitoptimization,
      title={Exact Quantum Circuit Optimization is co-NQP-hard}, 
      author={Adam Husted Kjelstrøm and Andreas Pavlogiannis and Jaco van de Pol},
      year={2025},
      eprint={2510.16420},
      archivePrefix={arXiv},
      primaryClass={quant-ph},
      url={https://arxiv.org/abs/2510.16420}, 
}

@article{McArdle2020,
  title = {Quantum computational chemistry},
  author = {McArdle, Sam and Endo, Suguru and Aspuru-Guzik, Al\'an and Benjamin, Simon C. and Yuan, Xiao},
  journal = {Rev. Mod. Phys.},
  volume = {92},
  issue = {1},
  pages = {015003},
  numpages = {51},
  year = {2020},
  month = {Mar},
  publisher = {American Physical Society},
  doi = {10.1103/RevModPhys.92.015003},
  url = {https://link.aps.org/doi/10.1103/RevModPhys.92.015003}
}

@article{clinton2024towards,
  title={Towards near-term quantum simulation of materials},
  author={Clinton, Laura and Cubitt, Toby and Flynn, Brian and Gambetta, Filippo Maria and Klassen, Joel and Montanaro, Ashley and Piddock, Stephen and Santos, Raul A and Sheridan, Evan},
  journal={Nature Communications},
  volume={15},
  number={1},
  pages={211},
  year={2024},
  publisher={Nature Publishing Group UK London}
}

@article{Georgescu2014,
  title = {Quantum simulation},
  author = {Georgescu, I. M. and Ashhab, S. and Nori, Franco},
  journal = {Rev. Mod. Phys.},
  volume = {86},
  issue = {1},
  pages = {153--185},
  numpages = {33},
  year = {2014},
  month = {Mar},
  publisher = {American Physical Society},
  doi = {10.1103/RevModPhys.86.153},
  url = {https://link.aps.org/doi/10.1103/RevModPhys.86.153}
}

@misc{Mikel2025Bounds,
  title = {Tight bound for the total time in digital-analog quantum computation},
  author = {Garcia de Andoin, Mikel and Sanz, Mikel},
  eprint={2512.11619},
  archivePrefix={arXiv},
  primaryClass={quant-ph},
  url={https://arxiv.org/abs/2512.11619}, 
  year = {to be published}
 }

@article{Lamata2018,
   title={Digital-analog quantum simulations with superconducting circuits},
   volume={3},
   ISSN={2374-6149},
   url={http://dx.doi.org/10.1080/23746149.2018.1457981},
   DOI={10.1080/23746149.2018.1457981},
   number={1},
   journal={Advances in Physics: X},
   publisher={Informa UK Limited},
   author={Lamata, Lucas and Parra-Rodriguez, Adrian and Sanz, Mikel and Solano, Enrique},
   year={2018},
   month=jan, pages={1457981} 
}

@article{abughanem2025superconducting,
  title={Superconducting quantum computers: who is leading the future?},
  author={AbuGhanem, Muhammad},
  journal={EPJ Quantum Technology},
  volume={12},
  number={1},
  pages={102},
  year={2025},
  publisher={Springer}
}

@article{arrazola2016digital,
  title={Digital-analog quantum simulation of spin models in trapped ions},
  author={Arrazola, I{\~n}igo and Pedernales, Julen S and Lamata, Lucas and Solano, Enrique},
  journal={Scientific reports},
  volume={6},
  number={1},
  pages={30534},
  year={2016},
  publisher={Nature Publishing Group UK London}
}

@article{bluvstein2024logical,
  title={Logical quantum processor based on reconfigurable atom arrays},
  author={Bluvstein, Dolev and Evered, Simon J and Geim, Alexandra A and Li, Sophie H and Zhou, Hengyun and Manovitz, Tom and Ebadi, Sepehr and Cain, Madelyn and Kalinowski, Marcin and Hangleiter, Dominik and others},
  journal={Nature},
  volume={626},
  number={7997},
  pages={58--65},
  year={2024},
  publisher={Nature Publishing Group UK London}
}

@article{Lu_2024,
   title={Digital–analog quantum learning on Rydberg atom arrays},
   volume={10},
   ISSN={2058-9565},
   url={http://dx.doi.org/10.1088/2058-9565/ad9177},
   DOI={10.1088/2058-9565/ad9177},
   number={1},
   journal={Quantum Science and Technology},
   publisher={IOP Publishing},
   author={Lu, Jonathan Z and Jiao, Lucy and Wolinski, Kristina and Kornjača, Milan and Hu, Hong-Ye and Cantu, Sergio and Liu, Fangli and Yelin, Susanne F and Wang, Sheng-Tao},
   year={2024},
   month=nov, pages={015038} }

@article{Choi2020,
  title = {Robust Dynamic Hamiltonian Engineering of Many-Body Spin Systems},
  author = {Choi, Joonhee and Zhou, Hengyun and Knowles, Helena S. and Landig, Renate and Choi, Soonwon and Lukin, Mikhail D.},
  journal = {Phys. Rev. X},
  volume = {10},
  issue = {3},
  pages = {031002},
  numpages = {27},
  year = {2020},
  month = {Jul},
  publisher = {American Physical Society},
  doi = {10.1103/PhysRevX.10.031002},
  url = {https://link.aps.org/doi/10.1103/PhysRevX.10.031002}
}

@article{bassler2025general,
  title = {General, Efficient, and Robust Hamiltonian Engineering},
  author = {Ba\ss{}ler, P. and Heinrich, M. and Kliesch, M.},
  journal = {PRX Quantum},
  volume = {6},
  issue = {4},
  pages = {040346},
  numpages = {33},
  year = {2025},
  month = {Nov},
  publisher = {American Physical Society},
  doi = {10.1103/9yxv-tdqr},
  url = {https://link.aps.org/doi/10.1103/9yxv-tdqr}
}

\end{document}